\documentclass[prl,final,twocolumn,superscriptaddress]{revtex4-1}
\usepackage{amsmath}
\usepackage{natbib}
\usepackage[normalem]{ulem}
\usepackage[pdftex]{graphicx}
\usepackage{epstopdf}
\usepackage{times}
\usepackage{txfonts}

\usepackage[colorlinks,bookmarks=false,citecolor=blue,linkcolor=red,urlcolor=blue]{hyperref}

\def\be {\begin{equation}}
\def\ee {\end{equation}}

\def\horparallel{ \lower.5ex\hbox{ \includegraphics[width=2ex]{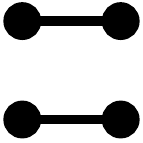}}\,\, }
\def\vertparallel{ \lower.5ex\hbox{ \includegraphics[width=2ex]{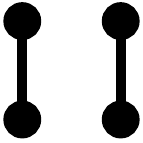}}\,\, }
\def\gsim{\mathrel{\rlap{\lower4pt\hbox{\hskip1pt$\sim$}}\raise1pt\hbox{$>$}}}
\def\lsim{\mathrel{\rlap{\lower4pt\hbox{\hskip1pt$\sim$}}\raise1pt\hbox{$<$}}}
\begin{document}

\title{Evidence of columnar order in the fully frustrated transverse field Ising model on the square lattice}
\author{Sandro Wenzel}
\affiliation{Institute of
  Theoretical Physics, \'Ecole Polytechnique F\'ed\'erale de Lausanne
  (EPFL), CH-1015 Lausanne, Switzerland}
\author{Tommaso Coletta}
\affiliation{Institute of
  Theoretical Physics, \'Ecole Polytechnique F\'ed\'erale de Lausanne
  (EPFL), CH-1015 Lausanne, Switzerland}
\author{Sergey E. Korshunov}
\affiliation{L. D. Landau Institute for Theoretical Physics RAS, 142432
Chernogolovka, Russia}
\author{Fr\'ed\'eric Mila}\email{frederic.mila@epfl.ch}
\affiliation{Institute of
  Theoretical Physics, \'Ecole Polytechnique F\'ed\'erale de Lausanne
  (EPFL), CH-1015 Lausanne, Switzerland}
\date{\today}

\begin{abstract}
Using extensive classical and quantum Monte Carlo simulations, we investigate the
ground-state phase diagram of the fully frustrated transverse field Ising model
on the square lattice. We show that pure columnar order develops in the
low-field phase above a surprisingly large length scale, below which
an effective $U(1)$ symmetry is present. The same conclusion applies to the Quantum
Dimer Model with purely kinetic energy, to which the model reduces in the zero-field
limit, as well as to the stacked classical version of the model. By contrast, the
2D classical version of the model is shown to develop plaquette order. Semiclassical
arguments show that the transition from plaquette to columnar order is a consequence
of quantum fluctuations.
\end{abstract}

\maketitle
The Ising model in a transverse field, introduced by De Gennes in the context
of ferroelectric systems \cite{deGennes1963} and defined by the Hamiltonian
\be
\label{eqn:HTIM}
H_\mathrm{TFIM}=\sum_{\langle i j \rangle} J_{ij} \sigma_i^z
\sigma_k^z - \Gamma \sum_j \sigma_j^x\,, \ee where $\sigma^\alpha$ are
Pauli matrices, $\Gamma$ the
transverse field and $J_{ij}$ Ising coupling constants, has
imposed itself as one of the most useful minimal models to study
quantum fluctuations \cite{book_chakrabarti}. In particular, the
simple ferromagnetic version of the model undergoes, when the
transverse field is increased, a phase transition to a paramagnetic
phase that embodies several of the basic aspects of quantum phase
transitions \cite{sachdevbook}. Frustrated versions of the model, i.e. versions in which it is impossible to minimize the energy for all bonds simultaneously, have been studied over the years \cite{liebmannbook}, with in general very rich phase diagrams \cite{book_chakrabarti}.

Recently, special attention has been payed to the subclass of Fully
Frustrated Transverse Field Ising Models (FFTFIM)
\cite*{MoessnerFFIM2000,MoessnerFFIMlong2001}. In these models, the
couplings are limited to nearest neighbors and have equal magnitudes,
but their signs are such that it is impossible to satisfy
simultaneously all bonds around any elementary plaquette, resulting in
a very strong ground state degeneracy. The AF Ising model on the
triangular lattice is a typical example. On the square lattice, this can
be achieved if the number of antiferromagnetic couplings around any elementary
plaquette is odd, resulting into an Ising model often referred to as
Villain's odd model \cite{VillainOddModel77}. All Ising models satisfying
this condition are equivalent, being related by a local gauge transformation.
In the present paper, we use the gauge with
antiferromagnetic (positive) couplings ($J_{ij}=J$) on every second
vertical line, and ferromagnetic (negative) couplings ($J_{ij}=-J$) for
all other bonds.

The interest in this family of models has raised because of their
connection to the Quantum Dimer Model (QDM) of Rokhsar and Kivelson
\cite{RKmodel88}.  First introduced in the context of the Resonating
Valence Bond (RVB) theory of high T$_c$ cuprate superconductors, the
QDM has become one of the paradigmatic models in the field of quantum
spin liquids after Moessner and Sondhi have shown that, on the
triangular lattice, it has an RVB phase \cite{MoessnerQDMTriang2001}. On the square lattice, the
QDM is defined by
\begin{align}
  \nonumber H_{\rm QDM}= \sum \Big[ & -t \left (\left| \horparallel
  \right\rangle\left\langle \vertparallel \right| + \rm{h.c.}\right) + \\ &
  \quad\quad V \left( \left| \horparallel \right\rangle\left\langle
  \horparallel\right| + \left| \vertparallel \right\rangle\left\langle
  \vertparallel \right|\,\right) \Big]\,,
  \label{rk-hamiltonian}
\end{align}
(in terms of hard core dimer objects $\vertparallel$) where $t$ and $V$ are the amplitudes of kinetic and potential terms, and
the sum runs over all elementary square plaquettes.
In the limit $\Gamma\rightarrow 0$, and up to a trivial degeneracy associated
to a global spin flip, the FFTFIM maps onto the
purely kinetic QDM with $V/t=0$ \cite*{MoessnerFFIM2000,MoessnerFFIMlong2001}.

Over the last few years, the properties of the QDM on several lattices
have been investigated in great details, and in most cases a consensus
has been reached regarding its phase diagram as a function of $V/t$
\cite{*[{For a recent review see, e.g., }][{}] MoessnerRamanReview},
with one noticeable exception: on the square lattice, while it is
clear that the QDM has a columnar phase for $V\to -\infty$ and a
staggered phase for $V/t>1$, the situation between these two limits is still controversial. It was suggested on the basis of
exact diagonalizations (ED) that the system has columnar order up to
$V/t=1$ \cite{SachdevQDM1989}, but subsequent investigations have
challenged this picture. ED on larger systems have been interpreted as
evidence of plaquette order in the range $(V/t)_c <V/t<1$ with
$(V/t)_c\simeq -0.2$ \cite{LeungQDM1996}.  The same conclusion has
been reached with Quantum Monte Carlo (QMC), with a critical ratio
$(V/t)_c\simeq 0.6$ significantly closer to $1$ though
\cite{SyljuasenQDM2006}. Finally, a new scenario \cite{RalkoMixedPhase2008}
has been put forward on the basis of ED, Green's function QMC, and a mapping
onto a height model \cite{HenleyHeightModel1997},
according to which a mixed
phase is present between $V/t\simeq 0$ and $V/t\simeq 0.8$, in which
all symmetries broken in the columnar and plaquette phases are
simultaneously broken. This scenario has
been supported by QMC on a related hard-core boson model
\cite{WesselMixedPhase08} and is not in direct conflict with Ref.~\cite{SyljuasenQDM2006} whose analysis was not designed to be sensitive to the presence of a mixed phase.

A major problem in studying the competition between columnar and plaquette order
in the QDM on the square lattice is that there is no simple limit where the
plaquette phase is known to exist for sure.
In this Letter, we investigate the same competition in the context of the FFTFIM.
The main advantage is that, as we shall see, the FFTFIM has a plaquette phase in
a simple limit, that of classical spins. A combination of large scale QMC simulations
and semiclassical arguments shows that the plaquette phase disappears when quantum fluctuations are introduced,
and that pure columnar order is stabilized from the paramagnetic phase down to zero transverse field, hence
also for the QDM for $V/t=0$, with no indication of a mixed phase.

\begin{figure}
\includegraphics[width=\columnwidth]{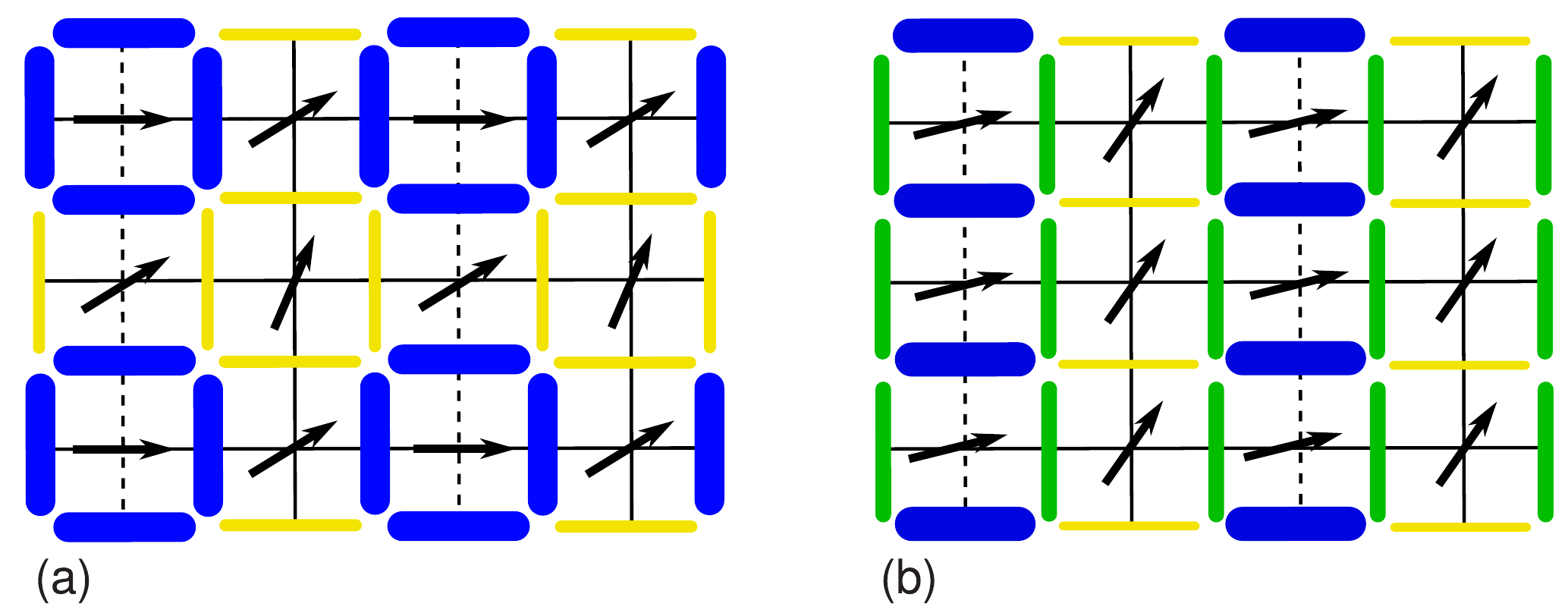}
\caption{\label{fig:dimerstructures}(color online). Visualization of
  the square lattice structure (black solid lines for ferromagnetic
  couplings and dashed lines for AF couplings) and of particular
  classical spin configurations with corresponding dimer
  configurations on the dual lattice: (a) plaquette dimer structure and (b) columnar dimer structure.
  Dimer densities on the dual lattice (encoded by the thickness of color lines) correspond to
  bond energies on the original spin lattice as described in the
  text.}
\end{figure}
A convenient gauge-invariant description of ordering in the
\makebox{FFTFIM} can be introduced \cite{Misguich2008,Colleta2011} by
using the energies of bonds $\langle ii+\mathbf{e}_\alpha\rangle$ of the
original lattice to define dimer densities
\makebox{$d_{i,\alpha}=\frac{1}{2} \left( 1 - J_{ii+\mathbf{e}_\alpha}/J
\sigma_i^z \sigma_{i+\mathbf{e}_\alpha}^z\right)$} on the bonds of the
dual lattice crossing them. This mapping gives rise to the visualization
of two typical dimer-solid states shown in Fig.~\ref{fig:dimerstructures}.
It allows one to characterize all aforementioned dimer-solid phases by a
single complex order parameter $\Psi$
\cite{SachdevQDM1989,LeungQDM1996,Jalabert91,Arnab07,WesselMixedPhase08},
which in terms of $\tilde d_{\alpha}(\mathbf{q})=\frac{1}{N} \sum_i
 e^{\mathrm{i}\, \mathbf{q}\cdot \mathbf{r}_{i,\alpha}}
 d_{i,\alpha}$, the Fourier transforms of dimer density
$d_{i,\alpha}$, has a very simple form,
\begin{equation}
 \label{eqn:Psi}
 \Psi=2[\tilde d_x(0,\pi)+ \mathrm{i}\,\tilde d_y(\pi,0)]\,.
 \end{equation}
Long-range ordering into some
sort of valence
 bond solid is indicated by $\langle |\Psi| \rangle > 0$, while
 the precise structure of the solid is controlled by the phase $\phi$ of the
 order parameter $\Psi = |\Psi|e^{\rm{i}\phi}$
 \cite{WesselMixedPhase08}: The columnar phase is
 characterized by a purely real or imaginary $\Psi$, i.e.,
 $\phi=\phi_c= n\pi / 2$ ($n=0,\ldots,3$), the plaquette state
 by a phase angle $\phi=\phi_p= \pi/4 + n\pi / 2$ ($n=0,\ldots,3$), 
 and the mixed-state by a phase $\phi$ which is neither of
 the form $\phi_c$ nor $\phi_p$.  Instead of directly studying the phase
 $\phi$, we find it more convenient to look at a phase sensitive order
 parameter \cite{IsakovFrustratedIsing2003,Lou2007} defined as
\begin{equation}
\label{eqn:fourfoldorder}
\Psi_4= \langle |\Psi| \cos(4\phi) \rangle.
\end{equation}
It is straightforward to see that in a columnar phase $\Psi_4\to
|\Psi|$ whereas $\Psi_4 \to -|\Psi|$ holds for plaquette ordering. In
a true mixed phase, however, $|\Psi_4|$ will not converge towards
$|\Psi|$ in the thermodynamic limit.

In addition, other studies \cite{RalkoMixedPhase2008, WesselMixedPhase08} used special order parameters
$P_+$ and $P_-$, based on plaquette structure factors \footnote{For each
  plaquette on the dual lattice, identified by a lattice site $i$ and
  position $\mathbf{r}_i$ on the original lattice, one considers
  symmetric/antisymmetric operators
  $k_{\pm,i}=d_{i,x}d_{i-\mathbf{e}_x,x}\pm
  d_{i,y}d_{i-\mathbf{e}_{y},y}$ and calculates the equal-time
  structure factor $S_{\pm}(\mathbf{q})=(1/N^2) \sum_{i,j}
  e^{\rm{i}\mathbf{q} (\mathbf{r}_i - \mathbf{r}_j)} k_{\pm,i}
  k_{\pm,j}$. Then $P_{-}=S_{-}(0,0)^{1/2}$ detects columnar
  correlations and $P_{+}=S_{+}(\pi,\pi)^{1/2}$ plaquette correlations.}, to measure plaquette and columnar ordering correlations separately. We will
show, however, that in the present context those have to be analyzed
with care.

\paragraph{Classical limit.---} The classical limit of Eq.~\eqref{eqn:HTIM} defined by
interpreting $\sigma^z$ and $\sigma^x$ as components of an ordinary
three-dimensional vector of length $S$ has previously been used to
study the FFTFIM on the honeycomb lattice \cite{Colleta2011}, which
can be mapped onto the QDM on the triangular lattice in the limit of
weak field, and its predictions, when combined with linear spin-wave
theory (LSWT), agree very well with QMC results on the QDM.
For the FFTFIM on the square lattice, the paramagnetic phase with
spins along the $x$ direction can be shown to extend down to
$\Gamma_c/J=2\sqrt{2}$, and in the range $0<\Gamma/J<2\sqrt{2}$, a single
phase is stabilized with a four-site unit cell and one spin per unit
cell fully polarized along the field. This has been checked
numerically and can be proven analytically in the limits
$\Gamma\rightarrow 0$ and $\Gamma\rightarrow \Gamma_c^-$
\cite{unpublished}.  This phase corresponds to a {\it plaquette} state
in the dimer language (see Fig.~\ref{fig:dimerstructures}(a)). The
columnar state would correspond to the dimer density pattern sketched
in Fig.~\ref{fig:dimerstructures}(b).  Minimizing the energy for a
unit cell which has the symmetry of the columnar state leads to a
configuration which is a saddle point of the total energy, but not a
local minimum, and \emph{a fortiori} not a global one.

\begin{figure}
\includegraphics[width=0.7\columnwidth]{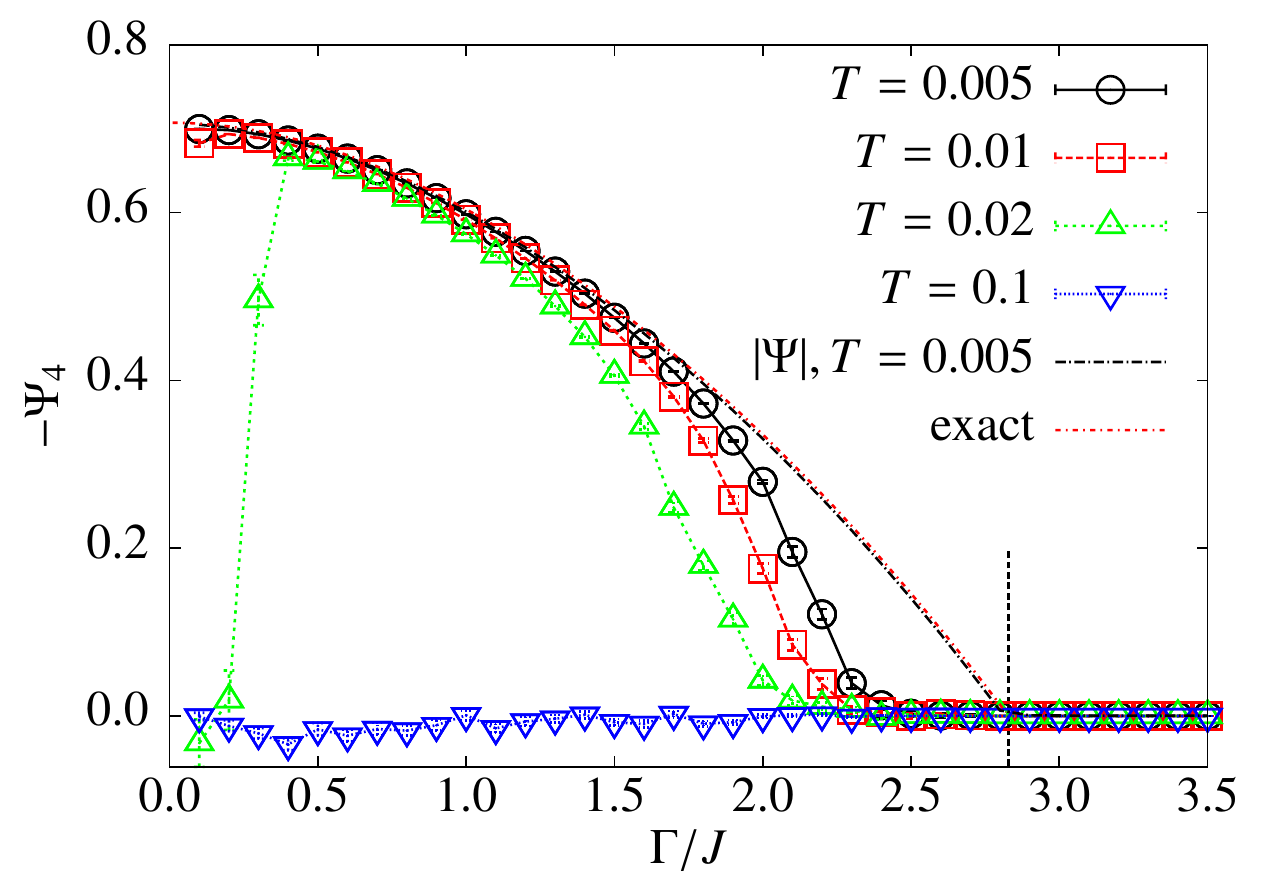}
\caption{\label{fig:PlaquetteClassical}(color online). Plot of the
  order parameter $-\Psi_4$ for the classical limit of
  \eqref{eqn:HTIM} at linear lattice size $L=16$ and various low
  temperatures $T$ confirming the nature of the ground state to be a
  plaquette state. For comparison, the norm $|\Psi|$ is shown at
  $T=0.005$ as well as calculated from the exactly known ground
  state. The exact zero-temperature critical field $\Gamma_c/J=2\sqrt{2}$ is
  indicated by the vertical line.}
\end{figure}

This result is quite surprising in view of the most recent results
obtained for the QDM at $V/t=0$, which suggest a columnar or a mixed
phase, but not a plaquette phase. In order to confirm it and to test
the order parameter $\Psi$ introduced above, we have performed finite temperature MC
simulations of the classical version of \eqref{eqn:HTIM} on systems
of size $L$ with periodic boundary conditions. They fully corroborate
our analysis, see Fig.~\ref{fig:PlaquetteClassical}. In particular, we
find a convergence of the phase sensitive quantity $-\Psi_4\to |\Psi|$
in the ordered region $\Gamma<\Gamma_c$, proving the presence of the
plaquette state in the classical FFTFIM.

\begin{figure}[t]
\includegraphics[width=\columnwidth]{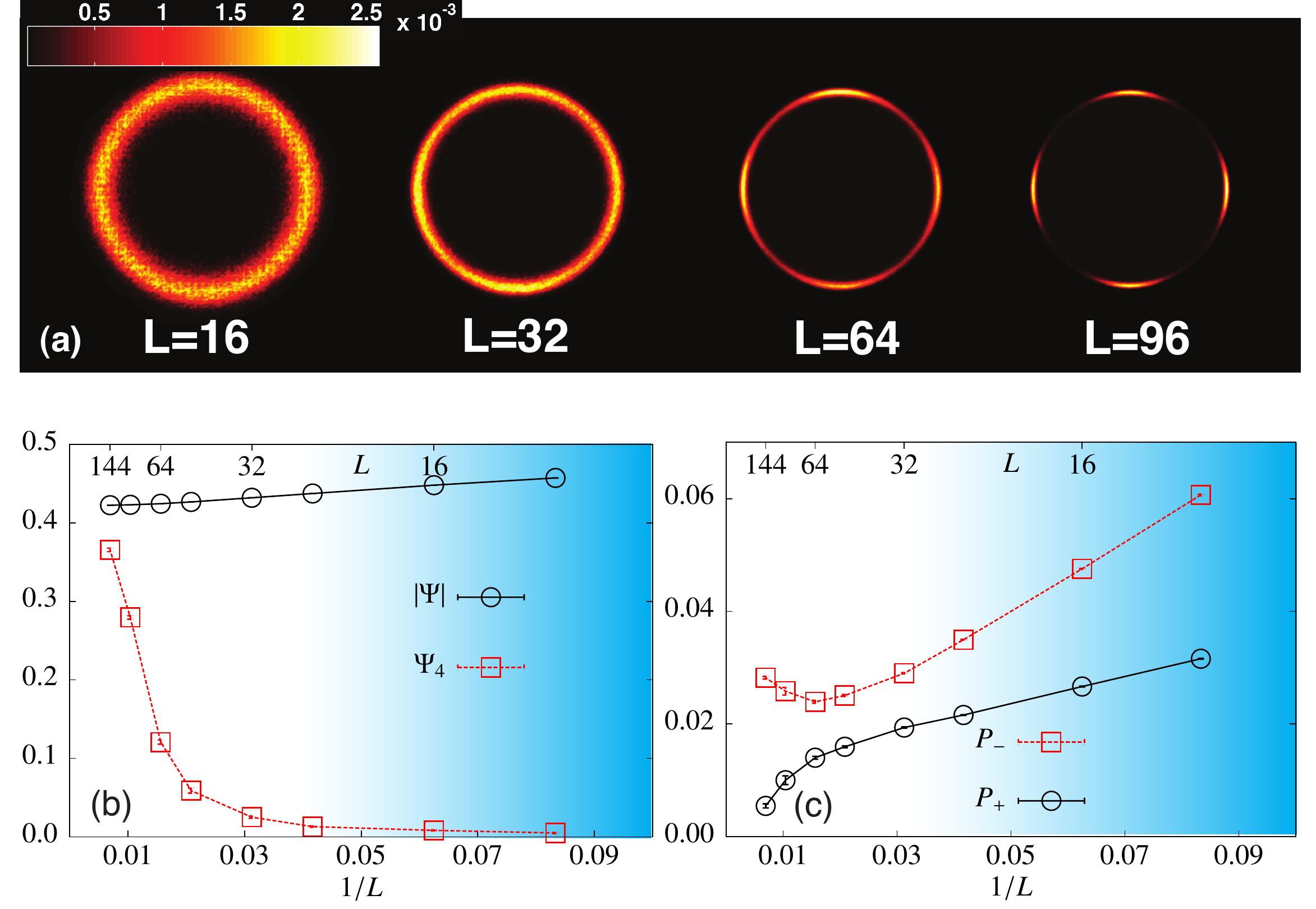}
\caption{\label{fig:stackedmagnet:scaling}(color online). (a)
  Order parameter distribution function (in the complex $\Psi$-plane)
  for the stacked magnet at $T=1.5$ for
  various lattice sizes. (b) Finite-size behavior of the two order parameters $\Psi$ and $\Psi_4$
  versus $1/L$. (c) Same as (b) for the order parameters $P_+$
  (plaquette correlations) and $P_-$ (columnar correlations). The color shaded regions indicate the
  approximate range where almost perfect $U(1)$-symmetric
  distributions are found. All results demonstrate
  the presence of columnar dimer solid in the stacked magnet.}
\end{figure}

\paragraph{The stacked Ising model.---} Motivated by this somehow unexpected result,
and as a first step towards a full quantum treatment of the FFTFIM, we
have decided to use the same type of MC simulations to revisit the
stacked fully frustrated Ising model first studied by Jalabert and
Sachdev \cite{Jalabert91}. This model is closely related to the FFTFIM
since they can be mapped onto each other using Trotter-Suzuki's
decomposition if appropriately scaled coupling constants and stack
dimension are used, and Jalabert and Sachdev have concluded that
columnar order is stabilized in the stacked model when the temperature $T$ is finite. However, what they actually showed is
that there is a phase transition at $T\simeq 2.85$ below which
$|\Psi|$ acquires a finite value without actually resolving the phase
of the order parameter. We have performed high-statistics
MC simulations inside the ordered phase at $T=1.5$, with a focus on
obtaining such conclusive phase-sensitive information. The resulting
distribution functions of the complex order parameter $\Psi$ for
linear lattice sizes $L=16,32,64,96$ are shown in
Fig.~\ref{fig:stackedmagnet:scaling}(a).  Interestingly, a $U(1)$
symmetric distribution is found for $L\lsim 32$, but ultimately four
peaks develop at phase angles $\phi=\phi_c$ and a clear \emph{columnar
  state} emerges above the $U(1)$ regime. These results lead to two
important conclusions: (i) Quantum fluctuations seem to change the
nature of ordering from plaquette to columnar; (ii) A $U(1)$ length
scale (as in \cite{Lou2007}) prohibits to draw any conclusion from
data on small system
sizes.

It is instructive to study the effect of this length scale on the finite-size
scaling of order
parameters. Figure~\ref{fig:stackedmagnet:scaling}(b,c) compares the
finite-size dependence of $|\Psi|$ and $\Psi_4$. It is only for
$L\gsim 32$ that a clear signal appears showing that $\Psi_4$ will
eventually converge towards $|\Psi|$.
For the order parameters $P_+$ and $P_-$, the effect is even more dramatic
(see Fig.~\ref{fig:stackedmagnet:scaling}(c)): The results for sizes up to $L=40$ are
clearly consistent with a finite value for both $P_+$ and $P_-$ in the thermodynamic
limit, hence with a mixed state, as in the analysis of Refs.~\cite{RalkoMixedPhase2008,WesselMixedPhase08},
but larger sizes reveal that $P_+$ finally drops to zero while $P_-$
starts increasing again to reach a much larger value than expected, a clear indication of a
pure columnar state.

\begin{figure}
\includegraphics[width=\columnwidth]{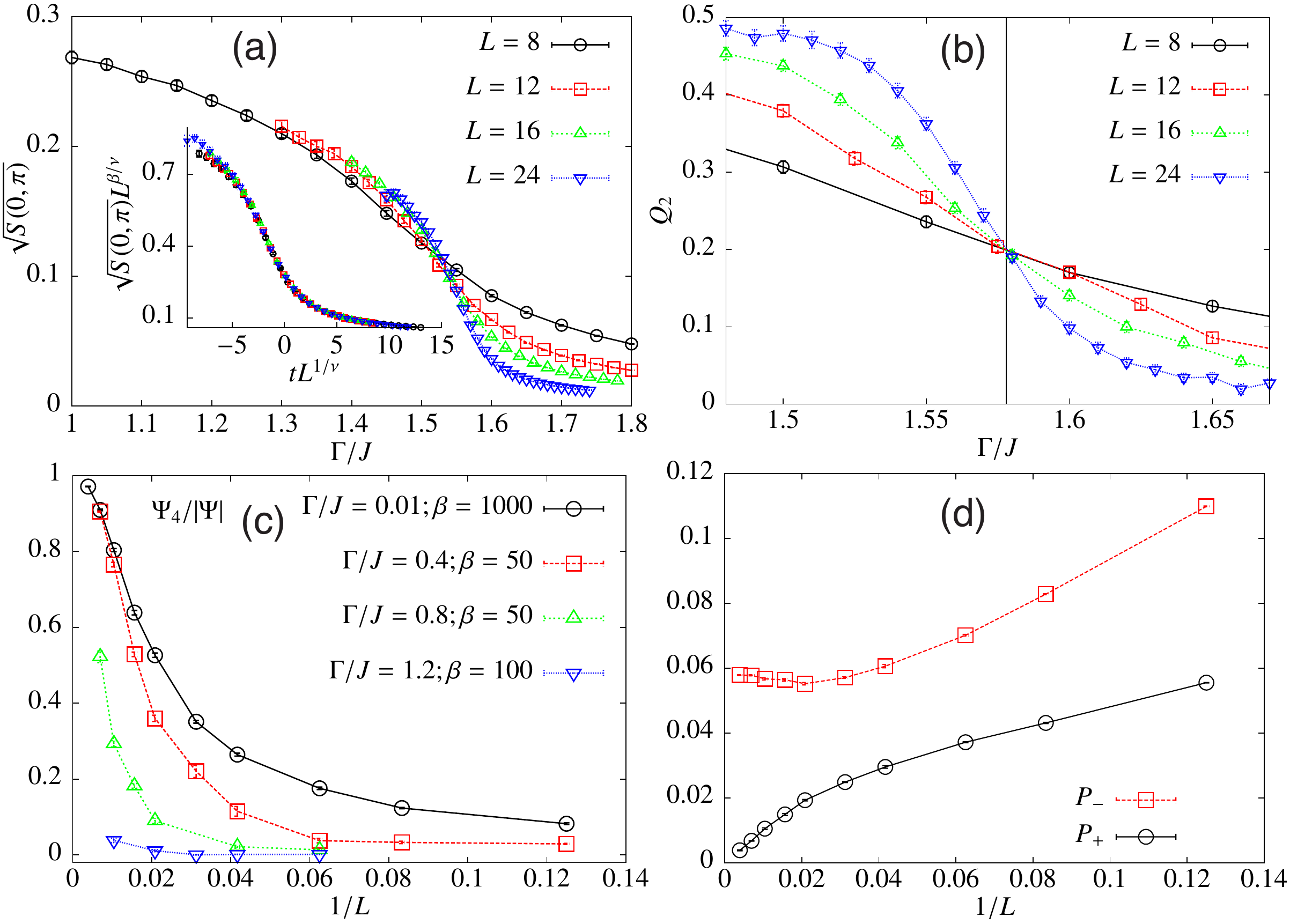}
\caption{\label{fig:quantum}(color online). (a) Spin structure factor
  $S(0,\pi)$ as a function of the transverse field $\Gamma$ indicating
  an ordered phase below $\Gamma/J \lsim 1.6$ (Inset : Scaling of
  $S(0,\pi)$ with the critical exponents from the $XY$ universality
  class.) (b) Associated Binder parameter. It gives a precise estimate
  of $\Gamma_c/J$ indicated by the vertical line. (c) Scaling of the
  order parameter ratio $\Psi_4/\Psi$ for the quantum FFTFIM for
  various values of the field $\Gamma$. It provides convincing
  evidence of columnar ordering for the smallest fields. (d) Analysis
  of the order parameters $P_+$ and $P_-$ for the smallest field studied
  $\Gamma/J=0.01$.}
\end{figure}

\paragraph{The quantum FFTFIM.---}
Next, we turn to a direct analysis of the ground-state phase
diagram of Eq.~\eqref{eqn:HTIM} for arbitrary field $\Gamma$. Using
continuous time and sign-problem free quantum Monte Carlo methods
\cite{RiegerQMC99,PhysRevLett.101.210602} on periodic lattices with
linear system sizes up to $L=256$, we investigate the order parameter
$\Psi$ and the basic ordering modes in terms of the (sublattice)
structure factors $S(0,0)$ and $S(0,\pi)$ (see,
Fig.~\ref{fig:dimerstructures}(b)). In each case, the inverse
temperature $\beta$ was carefully chosen to be large enough to sample
ground-state properties. Results for the spin structure factors are
reproduced in Fig.~\ref{fig:quantum}(a), which
show that an ordered phase is stabilized by quantum fluctuations
below $\Gamma/J\lsim 1.6$. A more precise estimate of the critical
transverse field is obtained from the Binder parameter $Q_2=1-\langle
S(0,\pi)^2 \rangle/(3 \langle S(0,\pi)\rangle^2)$ which is scale
invariant at the quantum critical point. The crossings of $Q_2$ lead
to the estimate $\Gamma_c/J=1.578(2)$, see
Fig.~\ref{fig:quantum}(b). This critical field is considerably reduced
as compared to the classical value $2\sqrt{2}$. Using the value obtained
for $\Gamma_c$ and 3D $XY$ critical exponents ($\beta$ and
$\nu$) \cite{CampostriniXY2006}, a good data collapse of the structure
factors is obtained, \emph{directly demonstrating} the 3D $XY$
universality class of the quantum critical point. This result
coherently adds to the indirect analysis performed in
Ref.~\cite{IsakovFrustratedIsing2003} and supports the field-theoretic
picture \cite{BlankschteinTriangular1984,BlankschteinSquare1984}
beyond reasonable doubt. It also confirms the similarity between the
triangular and square lattice FFTFIM as far as the critical behavior
is concerned.  An analysis of the FFTFIM at finite temperature that
reveals the presence of an extended critical phase will be presented
elsewhere \cite{unpublished}.  To determine the nature of
the phase(s) for $\Gamma<\Gamma_c$, we calculate the order parameter $\Psi_4$ and analyze its finite-size
scaling with respect to $|\Psi|$. Figure \ref{fig:quantum}(c) shows
the essential results of this analysis for a couple of selected field
values inside the ordered region. While it is clearly difficult to get
phase sensitive information close to the quantum critical point
(for $\Gamma/J\simeq 1.2$ and above) -- continuous distributions as in
Fig.~\ref{fig:stackedmagnet:scaling} (top) are seen up to the largest
system size -- we note that it actually becomes easier to obtain
structural information in the limit $\Gamma/J\to 0$. Indeed, for
$\Gamma/J=0.01$ the scaling of $\Psi_4$ is sufficient to conclude on the
presence of a columnar state, a result fully supported by the behavior
of $P_+$ and $P_-$, see Fig.~\ref{fig:quantum}(d).  Based on this
analysis, we can safely conclude that the columnar state is stabilized
at least up to $\Gamma/J\approx 1.0$. Above this value, it is currently
difficult to analyse the nature of the phase by means of
finite-lattice simulations. This particular difficulty close to the
quantum critical point is actually not surprising since the 3D XY
quantum critical point has a $U(1)$ symmetry which is at the origin of
the disturbing crossover behavior found here \cite{Lou2007}. A similar
observation has already been made in the case of the triangular
lattice \cite{IsakovFrustratedIsing2003}.

\begin{figure}
\includegraphics[width=0.8\columnwidth]{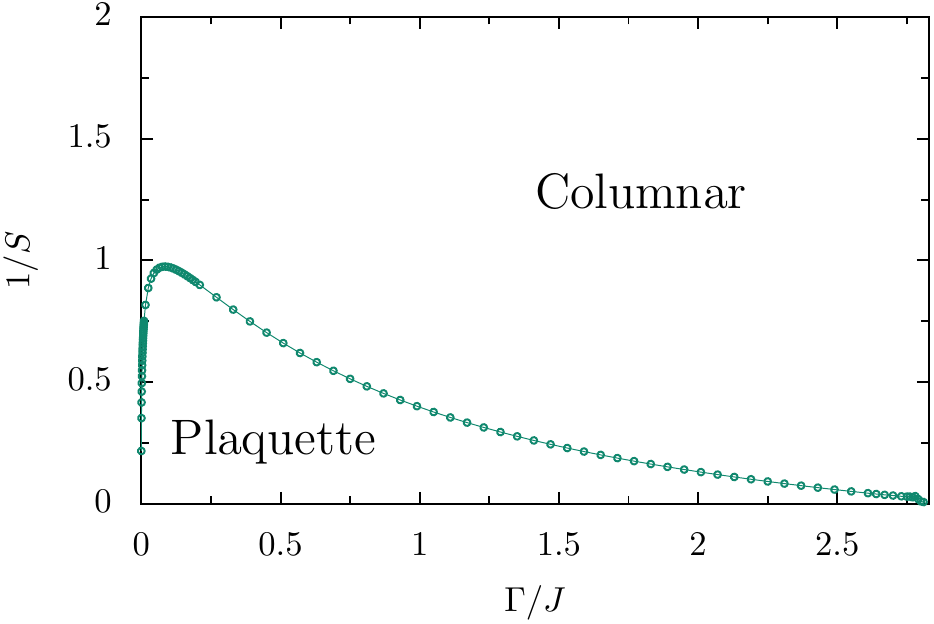}
\caption{\label{fig:semicl}(color online). Phase diagram in the
  $\Gamma/J-1/S$ plane, showing the phase boundary between columnar and
  plaquette phase where the transition line shown is an
    upper bound for the region of the stability of the plaquette phase. The columnar phase is clearly stabilized by
  quantum fluctuations.}
\end{figure}

\paragraph{Semiclassical approach.---}
All these results indicate that the order in the low field phase
changes from plaquette to columnar upon going from the classical to
the ultra-quantum spin-$1/2$ case.  In order to investigate this issue
from a complementary angle, we have studied how quantum fluctuations
may favor columnar order in the context of a semiclassical
approach. As stated earlier, the classical solution
which corresponds to columnar order is not a
local minimum of the energy. So the spectrum of the harmonic
fluctuations cannot be
diagonalized by a Bogoliubov transformation.
An upper bound of the zero point energy can nevertheless be obtained by applying local fields in the
direction of the spins to achieve the positiveness of the quadratic form that can
be diagonalized by a Bogoliubov transformation (see Supplemental
Material).  For the plaquette state, the Hamiltonian of the
quadratic fluctuations can be diagonalized by a Bogoliubov
transformation.  The phase diagram obtained by comparing the energy of
the columnar and plaquette states corrected in this way by quantum
fluctuations is shown in Fig.~\ref{fig:semicl}. The results nicely
confirm the anticipated transition: upon reducing the spin, quantum
fluctuations reduce the energy of the columnar state much faster than that
of the plaquette one and stabilize it for all fields for spin-$1/2$.

\paragraph{Conclusions.---} From the various results reported in this
paper, a number of firmly established conclusions can be drawn. First
of all, at low field, the ground state of the quantum FFTFIM on the
square lattice is an ordered phase whose structure corresponds, 
in the gauge invariant dimer language, to columnar order. This order is
stable at least up to $\Gamma/J=1$, and possibly up to the transition
to the paramagnetic phase, which has been shown to take place at
$\Gamma_c/J\simeq 1.58$. By contrast, the classical
version of the model develops an order with a four-spin unit cell at
low temperature which, in the dimer language, corresponds to plaquette
order.  Furthermore, the robustness of plaquette order with respect to
two types of perturbations, stacking and quantum fluctuations, has
been tested, with the conclusion that it is relatively fragile: 3D
stacking clearly stabilizes columnar order, while quantum fluctuations
treated within linear spin-wave theory stabilize columnar order in a
large portion of the $(1/S,\Gamma/J)$ phase diagram, and in particular
for all values of $\Gamma/J$ outside the paramagnetic phase for the
quantum model with $S=1/2$. 

Regarding the closely related QDM on the square lattice, the present
results clearly demonstrate that, in the purely kinetic regime
$V/t=0$, which is equivalent to the $\Gamma/J\rightarrow 0$ limit of
the FFTFIM, columnar order is stabilized.  In addition, this is likely
to be the case in a finite parameter range around this point since we
have not seen any particular indication of a proximity to a critical
point. Complementary work carried out in parallel to ours comes to
 similar conclusions \cite{Laeuchliprivate}.

Finally, the other important message of this paper is the presence of
a surprisingly large length scale below which the model has an
effective U(1) symmetry. This makes the identification of the actual
order stabilized in the thermodynamic limit particularly difficult
and, to a large extent, explains the conflicting results obtained so
far on the type of order stabilized in the QDM between $V/t=0$ and
$V/t=1$. By studying very large systems, we have been able to beat
this length scale up to $\Gamma/J\simeq 1$.  Beyond that value, the
proximity of the transition to the paramagnetic phase makes this
length scale even larger, and whether columnar order remains stable up
to the paramagnetic phase, as suggested by the semiclassical
approximation, could not be numerically decided.

\acknowledgements We thank P. Corboz, S. Isakov, A. L\"auchli, and S.~Wessel for
fruitful discussions and C. Henley for comments on the manuscript. This work was supported by the Swiss National
Fund and by MaNEP.

\bibliography{literature}

\appendix
\section{Supplemental Material}
Here we present the details of the analysis of the effect of quantum
fluctuations for the plaquette and columnar states. Quantum fluctuations
are investigated in the context of a large $S$ expansion (spin wave
expansion). In this framework, the fluctuations around a classical
solution are described in terms of Holstein-Primakoff bosons and the
fluctuation Hamiltonian is truncated at the harmonic level (linear spin
wave expansion).

In the two classical structures that are compared, the fluctuation
Hamiltonians do not contain terms which are linear in the
Holstein-Primakoff bosons and therefore are purely quadratic. In the case
of the plaquette state, the correction to the classical energy is obtained
by diagonalizing the fluctuation Hamiltonian. This is done in Fourier
space via a standard Bogoliubov transformation which yields the dispersion
relations and the correction to the classical energy.

The computation in the case of the columnar state is more involved. This
time the starting classical configuration is not a minimum but a saddle
point of the classical energy. This results in a quadratic fluctuation
Hamiltonian which is not positive definite. Trying to diagonalize the
quadratic Hamiltonian via a Bogoliubov transformation in this case yields
a spectrum which is not well defined for all values of momenta. For the
regions in the Brillouin zone near $\vec{k}=(0,\pm \pi)$ the spectrum is
not real but takes complex nonphysical values. In this case, the
correction to the classical energy cannot be computed. The fact that the
spectrum is not well defined results from the truncation of the spin wave
approximation to harmonic level. If this state is to become the ground
state when quantum fluctuations are included, higher order terms in the
spin wave expansion must yield a spectrum with real frequencies. However
this implies treating higher order terms self consistently.

In order to have a well-defined spectrum we proceed in a different
way. We add to the Hamiltonian a local field term on each site
of the form
\begin{equation}\nonumber
V=\frac{\delta}{S}\sum_j(S-\hat{S}_{j}^{z^\prime(j)}),
\end{equation}
where $\delta>0$ parametrizes the local field strength
and where the summation is taken over all sites and at each site
the axis $z^\prime(j)$ is directed along the direction of the classical
spin. The field $\delta$ is adjusted to have a
fluctuation Hamiltonian which is positive definite allowing it to be
diagonalized by a Bogoliubov transformation.
The resulting spectrum has real and positive frequencies with soft modes only at the wave
vectors $\vec{k}=(0,\pm \pi)$.
The advantages of this approach are the following: the addition of $V$ to the
Hamiltonian does not change the classical energy of
the state considered and allows to obtain dispersion relations which are
physically meaningful. Furthermore $V$ is
a strictly positive contribution to the Hamiltonian. Hence the corrections to
the energy of the columnar state computed with this approach provide an {\it
upper} bound of the energy of this state at order $1/S$.

We find that as a function of $S$ and $\Gamma/J$, the {\it upper} bound
of the energy of the columnar state is lower than the energy of the
plaquette state in a significant parameter range (see Fig.~5 of main text).
In particular, the columnar state wins for all fields for $S = 1/2$,
the value of the spin for which the model can be exactly mapped
onto a QDM with $V/t = 0$ in the limit $\Gamma/J\rightarrow 0$.

\end{document}